\documentstyle[emulateapj,psfig]{article}
\makeatletter

\newenvironment{inlinefigure}{%
\def\@captype{figure}%
\noindent\begin{minipage}{0.999\linewidth}\begin{center}}
{\end{center}\end{minipage}\smallskip}
\makeatother

\lefthead{Cowie et al.}

\begin{document}
\title{Detecting high redshift evolved galaxies as the hosts of optically faint
hard X-ray sources\altaffilmark{1}}
\author{L.\,L.\ Cowie,$\!$\altaffilmark{2,3}
A.\,J.\ Barger,$\!$\altaffilmark{2,3,4,5}
M.\,W.\ Bautz,$\!$\altaffilmark{6}
P.\, Capak,$\!$\altaffilmark{2}
C.\,S.\ Crawford,$\!$\altaffilmark{7}
A.\,C.\ Fabian,$\!$\altaffilmark{7}
E.\,M.\ Hu,$\!$\altaffilmark{2,3}
F.\,Iwamuro,$\!$\altaffilmark{8}
J.-P.\ Kneib,$\!$\altaffilmark{9}
T.\,Maihara,$\!$\altaffilmark{8}
K.\,Motohara$\!$\altaffilmark{8}
}

\altaffiltext{1}{Based in part on data collected at the Subaru Telescope,
which is operated by the National Astronomical Society of Japan}
\altaffiltext{2}{Institute for Astronomy, University of Hawaii,
2680 Woodlawn Drive, Honolulu, Hawaii 96822}
\altaffiltext{3}{Visiting Astronomer, W. M. Keck Observatory, jointly
operated by the California Institute of Technology, the University of
California, and the National Aeronautics and Space Administration}
\altaffiltext{4}{Hubble Fellow and Chandra Fellow at Large}
\altaffiltext{5}{Department of Astronomy, University of Wisconsin,
475 North Charter Street, Madison, WI 53706}
\altaffiltext{6}{Center for Space Research, 70 Vassar Street, Building 37,
Massachussetts Institute of Technology, Cambridge, MA 02139 }
\altaffiltext{7}{Institute of Astronomy, Madingley Road,
Cambridge CB3 0HA, U.K.}
\altaffiltext{8}{Department of Physics, Faculty of Science,
Kyoto University, Sakyo-ku, Kyoto 606-8502, Japan}
\altaffiltext{9}{Observatoire Midi-Pyr{\'e}n{\'e}es,
14 Avenue E. Belin, 31400 Toulouse, France}

\slugcomment{Astrophysical Journal Letters in press}

\begin{abstract}

We combine deep Subaru near-infrared images of the massive 
lensing clusters A2390 and A370 with Keck optical 
data to map the spectral energy distributions (SEDs) of 
{\it Chandra} X-ray sources lying behind the clusters. The three sources
behind A2390 are found to have extremely red colors
with SEDs consistent with
evolved galaxies at redshifts $z>1.4$. One source has
extremely anomalous colors, which we interpret as evidence for 
a type Sa SED at a redshift around 2.5.
The photometric redshift of another source has been
confirmed at $z=1.467$
from near-infrared spectroscopy using the CISCO spectrograph
on Subaru. Mapping of optically faint hard
X-ray sources may prove to be an extremely efficient way to
locate luminous evolved galaxies at high redshifts.

\end{abstract}


\section{Introduction}
\label{secintro}

Recent deep surveys with the {\it Chandra} X-ray Observatory 
have resolved most of the hard ($2-7$\ keV) X-ray background 
(XRB) into a population of point sources 
(\markcite{mushotzky00}Mushotzky et al.\ 2000;
\markcite{giacconi01}Giacconi et al.\ 2001), about
two-thirds of which are optically bright enough to be 
spectroscopically identified
(\markcite{barger01}Barger et al.\ 2001;
\markcite{horn01}Hornschemeier et al.\ 2001). 
A substantial fraction of the spectroscopically identified 
sources lie in the nuclei of luminous
bulge-dominated galaxies whose integrated optical spectra often
show little sign of the nuclear activity. 
The remaining third of the hard X-ray sources are optically 
faint ($I>23.5$) with $I-K$ colors that are very red. 
It has been speculated that
many of these optically faint sources may simply be higher redshift
analogs of the $z<1.5$ luminous galaxy hosts,
since the red SEDs would result in substantial optical fading
at higher redshift (\markcite{crawford00}Crawford et al.\ 2000;
\markcite{barger01}Barger et al.\ 2001).

In order to fully understand the origins of the hard XRB and 
to determine the evolution of the space density of AGN,
we need to know the nature and redshift 
distribution of the optically faint hard X-ray sources.
Furthermore, if it can be securely 
demonstrated that some of the optically faint hard X-ray sources 
do indeed lie in evolved galaxies at redshifts $z>1.5$, then the very 
existence of such systems will be of enormous interest 
in understanding the timescales for galaxy formation.
The combination of hard X-ray and optical data would
then provide an extremely efficient technique for locating
such sources.

Robust redshifts for evolved galaxies can be obtained from
their SEDs. Since much of the
signal for such an analysis lies in the relative slope
above and below the 4000 \AA\  break,
we need good infrared photometry for this purpose.

We have obtained deep near-infrared (NIR) imaging of the cores of 
two massive lensing clusters observed by {\it Chandra},
A2390 ($z=0.23$) and A370 ($z=0.37$),
to map the SEDs of the X-ray sources that lie behind the 
clusters. The advantage of targeting X-ray 
fields with foreground ($z=0.2-0.4$) massive cluster lenses 
is that any source lying behind will be substantially magnified.

\begin{figure*}
\vspace*{-0.5cm}
\centerline{\psfig{file=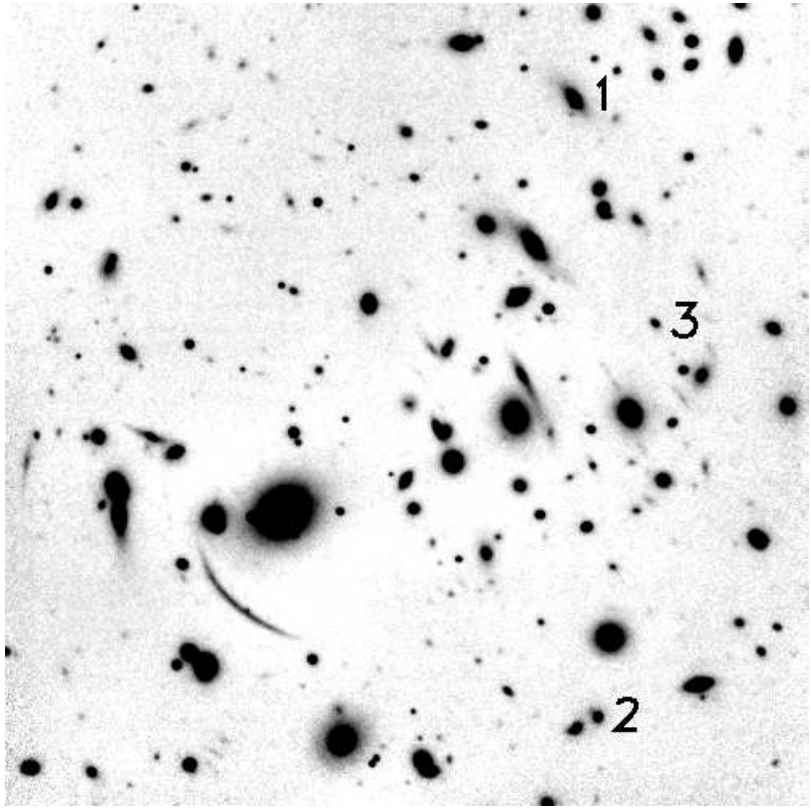,width=4.2in,angle=90}}
\caption{$K'$ image of the A2390 field with the three X-ray sources
marked. In each case the galaxy corresponding to the
X-ray source lies immediately to the left of the number.
The numbering corresponds to the listings in Table~\ref{tab1},
which is ordered by NIR magnitude. Coordinates (J2000) are:
(Source 1) RA: 21:53:34.03 Dec: 17:42:42.5, (Source 2) RA: 21:53:33.75
Dec: 17:41:16.8, (Source 3) RA: 21:53:33.23 Dec: 17:42:11.3.
}
\label{figimage}
\addtolength{\baselineskip}{-20pt}
\end{figure*}

The two hard X-ray sources detected behind the A370 
cluster by \markcite{bautz00}Bautz et al.\ (2000) are optically bright 
and have been spectroscopically identified ($z=1.060$ and $z=2.8$; 
\markcite{soucail99}Soucail et al.\ 1999;
\markcite{barger99}Barger et al.\ 1999;
\markcite{ivison98}Ivison et al.\ 1998). Their spectra show strong AGN
signatures, their morphologies are unusual, and both are strong
submillimeter sources (\markcite{bautz00}Bautz et al.\ 2000;
\markcite{smail97}Smail, Ivison \& Blain 1997). 
In contrast, the three X-ray sources detected behind the 
A2390 cluster by \markcite{fabian00}Fabian et al.\ (2000) are all 
coincident with relatively spatially smooth red galaxies
(Fig.~\ref{figimage}). Henceforth we refer to these as
Sources 1 (CXOUJ215334.0+174242), 
2 (CXOUJ215333.8+174116), and 
3 (CXOUJ215333.2+174211) (see Fig.~1).
These sources are not known submillimeter
sources and do not have optical spectroscopic 
redshifts (\markcite{crawford00}Crawford et al.\ 2000;
\markcite{fabian00}Fabian et al.\ 2000). 
While the faintest of these sources (Source 2)
is only detected in the soft band
of this still relatively shallow X-ray image, 
the three sources appear to be representative of the
optically faint hard X-ray source population.

The coordinates and properties of the three X-ray sources behind A2390 
are summarized in Table~\ref{tab1} and in the caption of Fig.~1.  
Magnitudes are measured in $2''$ diameter
apertures. We have applied a small differential correction relative to the
$K'$ magnitudes to allow
for differences in the image quality between the bands:
$-0.02 (H)$, $-0.04 (J)$, $0.03 (I)$, $-0.08 (R)$, $-0.18
(V)$, and $-0.17 (B)$. In the fitting procedures
we have also allowed for a 0.2 magnitude systematic uncertainty 
over the statistical errors in Table 1. 
Where total magnitudes are required, the offsets to
the isophotal magnitudes of the final column of the table should
be used.
The X-ray coordinates have been offset to match the optical data.
The Source 2 X-ray position has also been
updated from that given in \markcite{fabian00}Fabian et al.\ (2000) 
and now has a clear optical counterpart.

\section{Observations}
\label{secdata}

\subsection{Near-infrared Imaging}
\label{secnir}

We used the new Cooled Infrared Spectrograph and Camera for OHS
(CISCO, \markcite{moto98}Motohara et al.\ 1998) on
the Subaru 8.3\ m telescope UT 2000 June $18-19$, July $15-16$, 
September $10-12$, and November 7 to obtain extremely 
deep $J$, $H$, and $K'$
images of the clusters A2390 and A370. The detector
used was a $1024\times 1024$ HgCdTe HAWAII array with a pixel scale
of $0.111''$ for the June, July, and November runs and a
pixel scale of $0.105''$ for the 
September run. This provides a field-of-view
$\sim 2'\times 2'$. The data were taken in unguided mode and 
therefore relied on the superb telescope tracking to maintain image 
quality. To minimize the image degradation, a number of sub-exposures
were taken at each position 
in an eight-point pentagon pattern ($5''$ step size).
The weather conditions were photometric, and the seeing was 
typically $0.3''-0.5''$ during the first three observing runs, which 
was also the resolution for nearly all the final reduced images. Conditions
were clear but with variable seeing during the November
run, with characteristic image FWHM of $\sim0.8''$ for the A370 $H$ 
image taken on this night. 
The data were processed using median sky flats generated from
the dithered images. The data were calibrated from observations 
of the UKIRT faint standards (\markcite{irstds}Casali \& Hawarden 1992)
FS27, FS29, FS6, and FS10. The total exposure times for A2390 
were 5520\ s ($J$), 7290\ s ($H$), and 15360\ s ($K'$), and 
those for A370 were 13280\ s ($J$), 7680\ s ($H$), and 15360\ s ($K'$).
The $K'$ image of A2390 is shown in Fig.~\ref{figimage} with the three
X-ray sources marked.

\subsection{Optical Imaging}
\label{secopt}
Deep multicolor images of A370 were obtained using LRIS on the Keck
10\ m telescopes on UT 1999 August 11, September 9--10, and
2000 August 25.  The data were taken as a sequence of dithered exposures
with net integration times of 4200 s in $V\/$,
4800\ s in $R\/$, and 2700\ s in $I\/$. The data were
processed using median sky flats generated from the exposures.
Conditions were photometric during the observations.  The $V$, $R$,
and $I$ data were calibrated using the photometric and spectrophotometric
standard HZ4 (\markcite{turnshek90}Turnshek et al.\ 1990; 
\markcite{oke90}Oke 1990) and faint Landolt standard stars in the SA 95-42 
field (\markcite{landolt92}Landolt 1992). Deep $B$ (3780 s) 
and $R$ (2940s) images of A370
were obtained with ESI on Keck II on UT 2000 September 29--30.

For A2390, $B$ (3600\ s), $R$ (1800\ s), and $I$ (1080\ s) images
were obtained using ESI on the Keck II telescope on UT 2000 November 29--30.
The $V$ (4200\ s) image was obtained on UT 2000 September 29.
The data were calibrated with faint Landolt standards in the fields of
SA 113-337 and SA 95-42 (\markcite{landolt92}Landolt 1992).

\subsection{Near-infrared Spectroscopy}
\label{secspectra}

We used CISCO on UT 2000 November 7 with the
$zJ$ grating to obtain NIR spectra of two of the A2390 sources
(Sources 1 and 2).
The $zJ$ grating setup provides wavelength coverage
over the range $\lambda\lambda$8450--14100 \AA, with a steep decline
in sensitivity below $\sim8750$ \AA.  We used a $1''$ wide slit for the
observations, which provides a resolution of about R=280 over this wavelength
range, as measured from the FWHM of the neon calibration lines and 
spectral emission features in the targets. 
We took 400\ s exposures using a 2-point dither with $5''$ 
separation along the slit
at eight positions for Source 1, for a total of
3200\ s, and at ten positions for Source 2, 
for a total of 4000\ s.  A second order polynomial fit for the wavelength
was obtained from night sky lines and Paschen series lines in A type spectral 
template stars, and the zero-point was
adjusted to the position of OH sky lines in the target observations.

\begin{inlinefigure}
\centerline{\psfig{file=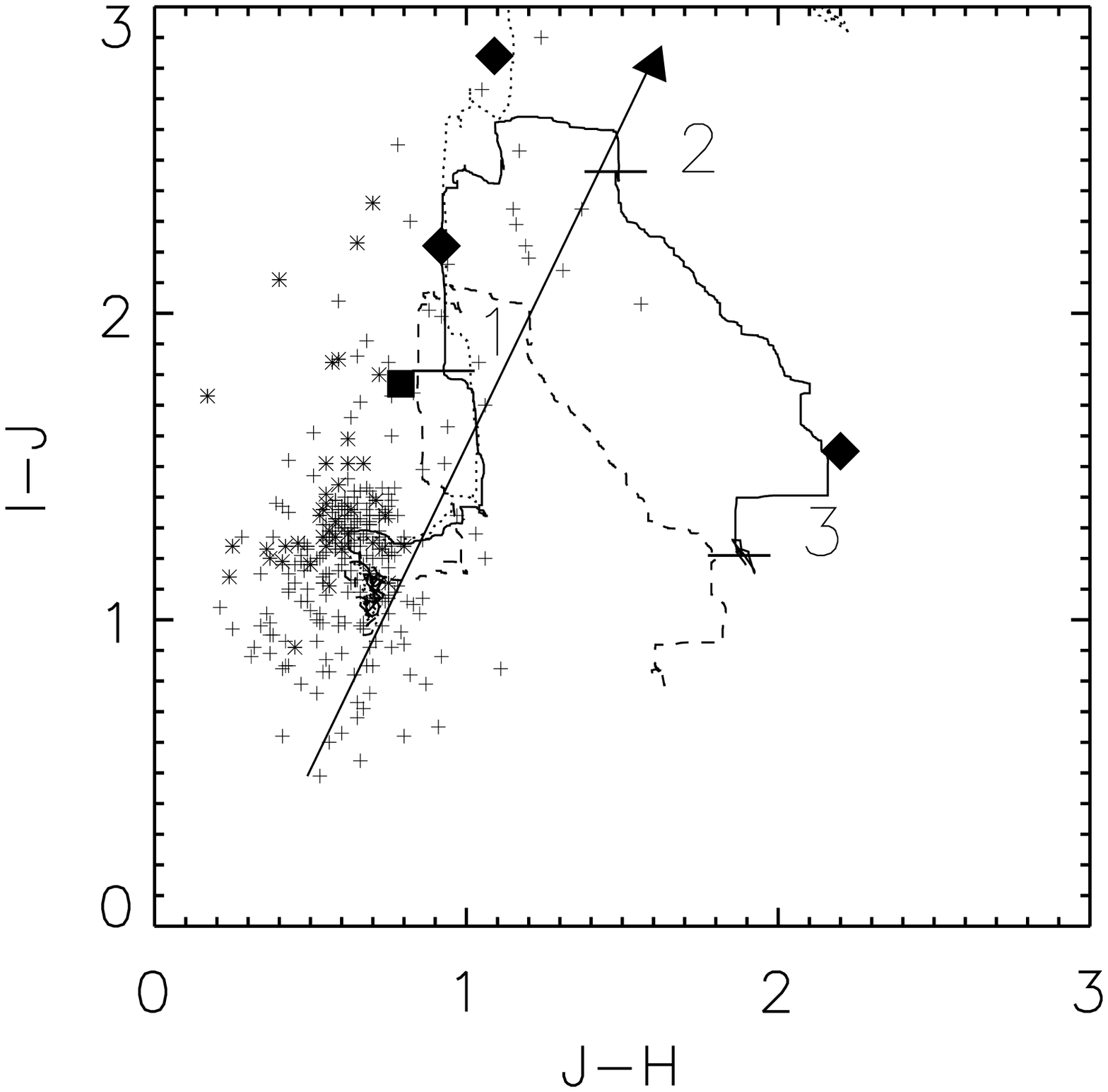,height=3.0in,angle=90}}
\vspace{6pt}
\caption{The $I-J$ versus $J-H$ colors for the sources in the A2390
and A370 CISCO fields are shown by crosses, with
compact sources shown as stars.
The four X-ray sources lying behind the clusters and
within the NIR area are
indicated by filled symbols, with the solid box showing the A370
source and the solid diamonds the A2390 sources. The elongated arrow shows
the locus of reddening stretching from an unreddened
flat spectrum source to one with $A_V=10$ magnitudes
of reddening. The solid curve shows the colors of
an Sa galaxy with redshift in the absence of any evolution, using the
local ($z=0$) SED from the Bruzual \& Charlot evolutionary code
(GISSEL98, Bruzual \& Charlot 1993).
The colors at redshifts $z=1$, 2, and 3 are
marked with ticks and labelled. The dotted line shows
the same colors for an elliptical galaxy (upper curve) and
the dashed line for an Sb galaxy (lower curve).
}
\label{figcolors}
\addtolength{\baselineskip}{10pt}
\end{inlinefigure}

\section{Results and Discussion}

\subsection{Colors and SEDs}

Figure~\ref{figcolors} shows the $I-J$ versus $J-H$ 
colors of sources detected in the 
A2390 and A370 CISCO fields. 
The four X-ray sources are indicated by filled symbols.
Only one of the A370 sources was covered by the infrared images, and it 
is shown by
the solid box. The three A2390 sources are shown as solid diamonds.
Point sources are shown as stars. All of the optical counterparts to the
X-ray sources are extended. Source 3 lies in 
an extremely unusual portion of the color-color plane. It has
a very large break beween the $J$ and $H$ bands and a much flatter $I-J$
color than the trend for the other galaxies. 
This portion of the plane cannot be reached by
reddening any normal galaxy or quasar spectrum. We
illustrate this in the figure where we show the locus of
a flat spectrum source with reddening as an arrow.
The colors are, however, consistent with a present-day Sa located at a
redshift near 3.
The remaining three X-ray sources, including the A370 galaxy known to be
at $z=1.060$, are consistent with being similar galaxies in 
the $z=1-2$ range.

\begin{inlinefigure}
\vspace*{0.03cm}
\centerline{\psfig{file=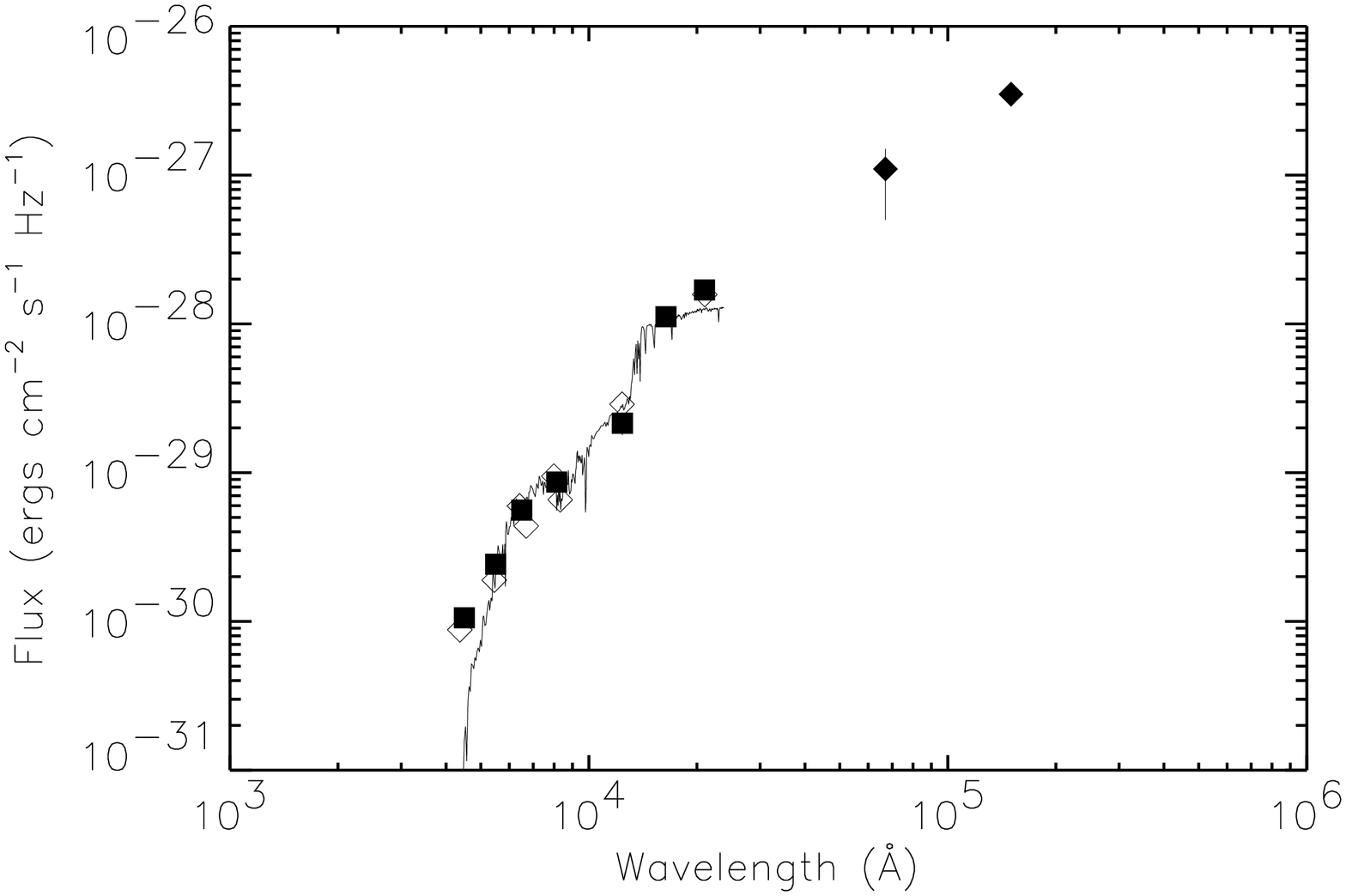,height=2.4in,angle=90}}
\caption{Observed SED for Source 3.
The $B$, $V$, $R$, $I$, $J$, $H$, and $K'$ fluxes from the present data
are shown as solid squares: the {\it HST} and ground-based data from
L{\'e}monon et al.\ (1998) are shown as open diamonds
together with their ISOCAM points (solid diamonds). Error
bars (where shown) are $1\sigma$.
The best-fit model from {\it hyperz} is overlaid.
}
\label{figseds}
\addtolength{\baselineskip}{-10pt}
\end{inlinefigure}

The red optical to NIR colors of these galaxies therefore appear to 
be a consequence of the SEDs rather than of dust reddening.
The morphologies also show no color differentiation.
This would not be expected if a reddened
nucleus were beginning to appear at the longer wavelengths. 
Photometric redshift fitting should be quite robust for this type 
of source.

\subsection{Photometric Redshifts}

We used the {\it hyperz} photometric fitting routine of
\markcite{hyperz}Bolzonella, Pell{\'o}, \& Miralles (2001)
and our observed $B$, $V$, $R$, $I$, $J$, $H$, and $K'$ magnitudes
to estimate redshifts for the X-ray sources.
The {\it hyperz} program compares the observed
SED of the galaxy to a set of template spectra
generated with the Bruzual \& Charlot evolutionary code (GISSEL98, 
\markcite{bc93}Bruzual \& Charlot 1993). We used the eight synthetic 
star formation histories constructed to match the observed properties of 
local field galaxies from type E to Im. We assumed no reddening.

Figure~\ref{figseds} shows the observed SED (solid squares)
of Source 3 with its deep break between the $J$ and $H$ bands.
The best-fit solution from {\it hyperz}
is overlaid on the observed SED. The photometric redshifts
for the three X-ray sources behind A2390
are $z_{phot}=1.4(1.3-1.6)$ (Source 1), 
$z_{phot}=1.5(1.3-1.7)$ (Source 2),
and $z_{phot}=2.6(2.4-2.7)$ (Source 3),
where the bracketed numbers are the 90\% confidence ranges. 
All of the SEDs are adequately fit by single burst models.
The youngest source (age $5\times 10^8$\ yr) is also the 
highest redshift source (Source 3).
While the photometric redshift estimates are not sensitive
to the details of the models, the age estimate is affected
by the choice of cosmology and the choice of models. For example,
a lower metallicity would require a larger age.
The source behind A370 is found to have
$z_{phot}=1.0(0.8-1.2)$, consistent with its spectroscopic 
redshift $z_{spec}=1.060$. The inclusion of reddening does not 
change the redshift estimates.

We note that \markcite{lemonon98}L{\'e}monon et al.\ (1998) and 
\markcite{wilman2000}Wilman, Fabian \& Gandhi (2000) put Source 3
at $z_{phot}\simeq 0.5-0.6$. 
The addition of the $H$ band data breaks the degeneracy in the
redshift information and places the source at a much higher
redshift in order to reproduce the strong $J-H$ break.

\begin{inlinefigure}
\vspace{-0.7cm}
\centerline{\psfig{file=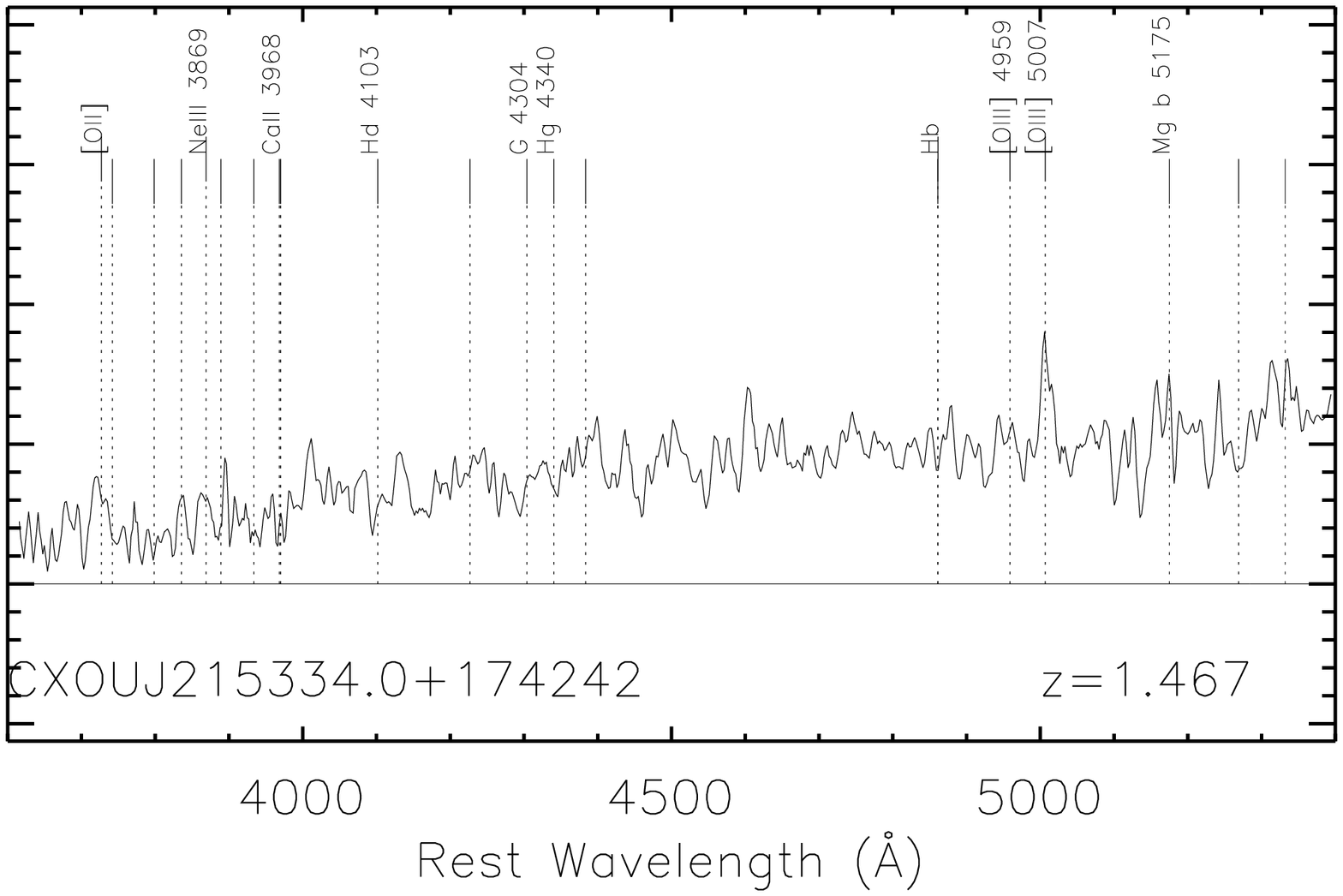,width=3.7in,angle=90}}
\vspace{-10pt}
\caption{CISCO NIR spectrum of Source 1
at $z_{spec}=1.467$. The [O$\,$II]$\,3727$
and [O$\,$III]$\,5007$ lines are clearly visible in the
spectrum, along with the Ca H and K absorption lines.
}
\label{figspectra}
\addtolength{\baselineskip}{-30pt}
\end{inlinefigure}

\subsection{Spectroscopic Redshifts}

Subsequent to estimating the photometric redshifts we obtained spectra 
of Sources 1 and 2 with CISCO. Source 1 (Fig.~\ref{figspectra})
has an unambiguous redshift of 1.467, based on the detection
of [OII]$\,3727$ and [OIII]$\,5007$. The weaker [OIII]$\,4959$ feature was too
faint to be detected. The [OII]$\,3727$ line was confirmed with
a one hour spectrum ($\lambda\lambda6000-10000$~\AA, 1.4$''$ wide slit,
$\sim12$~\AA\ resolution) taken with LRIS on Keck I on UT 4 October 2000.
This higher resolution spectrum yielded a redshift of 1.466.
The apparent weakness of the emission lines in the spectrum
is a consequence of the low resolution of the infrared spectroscopic
observations, and the equivalent widths of the lines are
quite representative of other field galaxies at these redshifts.

Source 2 has only a single line which, if interpreted as
[OIII]$\,5007$, would also place this source at 1.467, 
consistent with the photometric estimate. 
[OII]$\,3727$ is not clearly present, however, so it
is also possible that this line could be [OIII]$\,5007$, placing
the galaxy at $z_{spec}=2.317$. No LRIS spectrum is presently available 
for this source.

A one hour LRIS spectrum was also obtained for Source 3 
covering the 6000 to 10000~\AA\ region, but, as is reasonable given
the faintness of the source, no features could be clearly
seen.

\subsection{Longer wavelengths}

Based on ISOCAM data (\markcite{lemonon98}L{\'e}monon et al.\ 1998,
\markcite{altieri99}Altieri et al.\ 1999) all three of the galaxies
have significant excess flux in the mid-infrared wavelengths
over and above the galaxy light (see Fig.~\ref{figseds}), 
while none have strong
850 micron detections (\markcite{fabian00}Fabian et al.\ 2000;
\markcite{barger01b}Barger \& Kneib 2001).
This longer wavelength excess in the infrared and the absence of
strong submillimeter light is best interpreted with a
model of hot dust surrounding the central AGN, as discussed
in \markcite{wilman2000}Wilman, Fabian \& Gandhi (2000). 
We postpone a more detailed discussion to a subsequent paper.

\section{Conclusions}

The present observations, while based on a small
sample, appear to confirm that many of the
optically faint {\it Chandra} sources lie in luminous
high redshift galaxies which are relatively evolved.
All three X-ray sources behind A2390 lie in red galaxies with
redshifts between $z=1.4$ and $z=3$. The observed
absolute $K'$ magnitudes are [$-27.1,-26.5,-26.7$] for
sources [1,2,3], where Source 1 is placed at its
spectroscopic redshift and Sources 2 and 3 at their
photometric redshifts of $z_{phot}=1.5$ and $z_{phot}=2.5$.
When the cluster magnification of [2.1,2.8,7.8]
and the aperture to isophotal correction is accounted for, 
these become [$-27.2,-25.8,-24.8$],
where all absolute magnitudes are calculated
for $H_0$=65 and an $\Omega={1\over 3}$, $\Lambda={2\over 3}$ Universe.
Thus, the galaxies are comparable to or slightly
more luminous than the local $L_\ast$. When corrected
for magnification, the $2-7$\ keV X-ray luminosities lie in the
$10^{44}$ to $10^{45}$\ erg\ s$^{-1}$ range. All of
the sources have significant excess light in the
mid-infrared, presumably as a result of reprocessing of the
AGN light by hot dust 
(\markcite{wilman2000}Wilman, Fabian \& Gandhi 2000).

The very large magnification of the evolved high
redshift Source 3 is the key to its identification.
There may be numerous objects of this type in field
samples which are simply too faint to pick out using
NIR color selection. The
combination of very deep hard X-ray and optical data will
allow us to select such sources for intensive study. For
the highest redshift source, the demagnified X-ray flux
would be $3\times 10^{-15}$\ erg\ cm$^{-2}$\ s$^{-1}$ and the $I$
magnitude would be 26. It is at these levels that we may 
find further examples in the field samples.

\acknowledgements
AJB acknowledges support from NASA through Hubble
Fellowship grant HF-01117.01-A awarded by the
Space Telescope Science Institute, which is operated by the
Association of Universities for Research in Astronomy, Inc.,
for NASA under contract NAS 5-26555. AJB, LLC, and EMH acknowledge
support from NSF through grants AST-0084847, AST-0084816
and AST-0071208, and CSC and ACF thank the Royal Society for support.
We would like to thank Rolf Kudritzki
and Toni Songaila for critically reading the first draft
of this paper.


\begin{deluxetable}{cccccccccc}
\renewcommand\baselinestretch{1.0}
\tablewidth{0pt}
\tablecaption{X-ray Sources Behind A2390 Cluster}
\parskip=0.1cm
\tablenum{1}
\scriptsize
\tablehead{
\colhead{CXSOUJ} & Soft/Hard & $B$ & $V$ &
$R$ & $I$ & $J$ & $H$ & $K'$ & $K'_{iso}\tablenotemark{b}$ \cr
{Source Name} & {\hbox to 0pt{\hss{X-ray Fluxes\tablenotemark{a}}\hss}} & (mag) & (mag) & (mag) & (mag) & (mag) & (mag) & (mag) & (mag) \cr
}
\startdata
215334.0+174242 & 9.9/90 & $26.11\pm 0.40$ &
$25.39\pm 0.22$ & $23.93\pm 0.08$ & $22.17\pm 0.06$ &
$19.33\pm 0.01$ & $18.24\pm 0.00$ & $17.41\pm 0.01$ & 
$16.5$ \cr

215333.8+174116 & $3.7/<17$ & $24.32\pm 0.09$ &
$23.64\pm 0.04$ & $22.76\pm 0.03$ & $22.09\pm 0.06$ &
$19.87\pm 0.02$ & $18.95\pm 0.01$ & $18.15\pm 0.01$ &
$17.7$\cr

215333.2+174211 & 5.9/23 & $26.27\pm 0.45$ &
$25.25\pm 0.16$ & $24.23\pm 0.11$ & $23.67\pm 0.20$ &
$22.12\pm 0.14$ & $19.92\pm 0.02$ & $19.05\pm 0.03$ &
$18.8$ \cr

\enddata
\tablenotetext{a}{0.5-2.0 keV/2-7 keV X-ray fluxes given in units of
$10^{-15}$\ erg\ cm$^{-2}$\ s$^{-1}$}
\tablenotetext{b}{Isophotal magnitude to 2\% of peak surface brightness}
\label{tab1}
\end{deluxetable}

\end{document}